\documentclass[fleqn,a4paper,10pt]{article}
\usepackage[utf8]{inputenc}
\usepackage[pdftex]{graphicx}
\usepackage{amsmath}
\usepackage{amssymb}
\usepackage[hmargin=2.5cm,vmargin=3.5cm]{geometry}
\usepackage{color}
\usepackage[usenames,dvipsnames]{xcolor}
\usepackage{wasysym}
\usepackage{enumerate}
\usepackage{setspace}
\usepackage{placeins}
\date{}

\title{Voltage equilibration for reactive atomistic simulations of electrochemical processes}
\author{Nicolas Onofrio$^1$ and Alejandro Strachan$^1$\footnote{Corresponding author: strachan@purdue.edu}}

\begin{document}

\maketitle

\noindent
$^1$School of Materials Engineering and Birck Nanotechnology Center Purdue University, West Lafayette, IN 47906 USA \\

\begin{abstract}
We introduce EChemDID, a model to describe electrochemical driving force in reactive molecular dynamics simulations. 
The method describes the equilibration of external electrochemical potentials (voltage) within metallic 
structures and their effect on the self consistent partial atomic charges used in reactive molecular dynamics. 
An additional variable assigned to each atom denotes the local potential in its vicinity and we use fictitious, but 
computationally convenient, dynamics to describe its equilibration within not-simply connected metallic structures 
on-the-fly during the molecular dynamics simulation. This local electrostatic potential is used to dynamically modify 
the atomic electronegativities used to compute partial atomic changes via charge equilibration. 
Validation tests show that the method provides an accurate description of the electric fields generated by the applied 
voltage and the driving force for electrochemical reactions. 
We demonstrate EChemDID via simulations of the operation of electrochemical metallization cells. The simulations 
predict the switching of the device between a high-resistance to a low-resistance state as a conductive metallic bridge 
is formed and resistive currents that can be compared with experimental measurements. 
In addition to applications in nanoelectronics, EChemDID could be useful to model electrochemical energy conversion
devices. 

\end{abstract}


\section{Introduction}

Molecular dynamics~\cite{ISI:A1957WA17400039} (MD) modeling has become a key tool in many areas of science and engineering including 
chemistry,~\cite{ISI:A1987L587300003,KarplusNobel} material science~\cite{Holian:1998p538,Curtin2013} and biology~\cite{ISI:000177656200007}. 
Cloud computing has significantly simplified access to these MD simulations~\cite{StrachanCiSE} to the point where they can now be used in undergraduate 
education~\cite{brophy2013lectures}. While {\it ab initio} MD~\cite{ISI:A1985AUN4900027} (where atomic forces are computed from a quantum 
mechanical electronic structure calculation) plays an important role in many applications~\cite{ISI:A1996UW83600029,ISI:A1997WG77700056}, its 
computational intensity limits its use to relatively small systems and short times. 
Therefore, dynamical simulations with empirical interatomic potentials represent an indispensable tool to connect electronic processes with the response 
of materials or biological systems and in determining the atomistic mechanisms that govern their behavior. 
The last decades witnessed the development of accurate interatomic potentials for a wide range of materials and applications including 
metals~\cite{ISI:A1984SX73500005,ISI:A1992JG57300009}, semiconductors~\cite{ISI:A1985AGA1200057,ISI:A1988N126800045}, oxides
~\cite{ISI:A1985ACR0900010,streitz}, and molecular systems~\cite{ISI:A1990EQ50600010,ISI:A1996TY69800032,ISI:000267269600001}.
More recently, reactive potentials~\cite{ISI:000171614800010,ISI:A19633104B00020,ISI:A1994PH98700073} have demonstrated the ability to describe 
complex chemistry including the reaction and decomposition of high-energy density materials~\cite{ISI:000185235000056}, the description of solid/liquid 
interfaces~\cite{ISI:000277422900032,ISI:000281404400012} and the mechanisms of complex catalytic chemical reaction~\cite{ISI:000226449900011}.

A wide range of applications of technological interest require the description of electrochemical processes; examples include batteries~\cite{ISI:000172150700056},
capacitors~\cite{ISI:000260472800016}, and electrochemical metallization (ECM) cells for nanoelectronics~\cite{ISI:000250615400020}. 
Despite recent progress~\cite{ISI:000323177900006,ISI:000301984600017} we lack a self-consistent approach to describe how externally applied 
electrochemical potentials affect partial atomic charges and drive electrochemical reactions. 
In this paper we introduce the electrochemical dynamics with implicit degrees of freedom (EChemDID) method that describes the electrochemical
driving force resulting from the application of an external 
voltage to metallic electrodes in reactive MD simulations. A significant challenge addressed by our method is describing the equilibration of the
external potential over metallic electrodes with changing composition and morphology as ions electrochemically dissolve into the electrolyte or are deposited. 
We demonstrate the power of EChemDID via simulations of nanoscale ECM cells of interest for resistive switching non-volatile memory applications. 
The original version of the approach used a cluster analysis to determine the extent of the electrodes and enabled the first atomistic description of the operation 
of ECM cells \cite{NicoECM}. This paper extends and validates the method, provides a detailed description of the underlying physics and 
further exemplifies its use.

The remainder of the paper is organized as follows. Section 2 describes the EChemDID method and its implementation within reactive MD simulations.
Section 3 verifies our implementation and validates the model while Section 4 exemplifies its use to describe ECM cells of interest in nanoelectronics.
Finally, conclusions are drawn in Section 5.

\section{The EChemDID method}

This section starts with a brief review of reactive MD simulations focusing on the self-consistent calculation of partial atomic charges using charge 
equilibration (QEq~\cite{ISI:A1991FH09800070}). We then introduce the EChemDID approach to describe the application of an electrochemical potential 
to metallic electrodes. 

\subsection{Reactive MD and charge equilibration}

Reactive interatomic potentials like ReaxFF~\cite{ISI:000171614800010}, REBO~\cite{ISI:000086231600043,ISI:000174135200019} or 
COMB~\cite{ISI:A1994PH98700073}, use the concept of bond order to describe covalent interactions that include bond stretch, angle 
bending and torsional potentials as well as terms to penalize over and under coordination. In addition to covalent interactions, van der Waals interactions describe 
London dispersion and Pauli repulsion and electrostatics are described in terms of partial atomic charges. Theses charges need to be computed self-consistently based on 
the atomic structure of the system to capture charge transfer during chemical reactions. ReaxFF uses the charge equilibration method~\cite{ISI:A1991FH09800070}, also
known as  electronegativity equalization (EEM~\cite{ISI:A1986D312000013}), where the atomic charges are obtained equating the electronic chemical potential of every atoms in the system. The atomic electronegativity is obtained from a simple expression for the total electrostatic energy of the system:

\begin{equation}
E\left(\{q_i\};\{R_i\}\right)=\sum_i\left(\chi_i^0q_i+\frac{1}{2}H_iq_i^2\right)+\sum_{i<j}q_iq_jJ\left(|R_i-R_j|\right)
\label{eq:qeq}
\end{equation}

where the second term in the right-hand side (RHS) represents the electrostatic interaction between ions $i$ and $j$; $J$ takes the form of shielded Coulomb interaction. 
The first term represents the energy cost associated with changing the charge of individual atoms and depends on their electronegativity ($\chi_i^0$) and hardness ($H_i$). 
Given an atomic structure, the equilibrium partial atomic charges are obtained by equating the chemical potential of each atom:

\begin{equation}
\chi_i = \chi_0 = \frac{\partial E\left(\{q_i\};\{R_i\}\right)}{\partial q_i} = \chi_i^0+H_iq_i+\sum_{j\neq i}q_jJ\left(|R_i-R_j|\right)
\end{equation}

leading to $N$ equations with $N+1$ unknowns (charges for the $N$ atoms and the chemical potential $\chi_0$). The system of equations is closed by constraining the total
charge in the system: $Q=\sum_i q_i$. 

Important recent advances to charge equilibration methods include methods based on charge transfer between bonded atoms like the Split Charge Equilibration 
(SQE~\cite{ISI:000240351500011}), the addition of integer charge changes to describe reactions with ions dissolved in an electrolyte~\cite{ISI:000323177900006} 
and more recently, the extension of SQE and its derivation from density functional theory (DFT) as proposed in the atom-condensed Kohn-Sham DFT approximated 
to second order method (ACKS2~\cite{ISI:000315263500013}).
In this paper we develop EChemDID starting from the original formulation of QEq that remains widely used, the method is easily extensible to SQE.

We are interested in simulating electrochemical reactions that originate from the application of an external electrochemical potential difference ($\Phi_0$) between 
two electrodes; this has the effect of changing the relative energy of electrons in the two electrodes by $\Phi_0 e$. 
From the energy expression in Eq. \ref{eq:qeq} we see that this can be accomplished by changing the atomic electronegativity of the atoms in one electrode to 
$\chi_i^0\rightarrow\chi_i^0-\Phi_0/2$ and of those in the other electrode to $\chi_i^0\rightarrow\chi_i^0+\Phi_0/2$. Changes in atomistic electronegativity has been 
used in the past to create external fields and electrostatic potentials~\cite{ISI:000268613700015,ISI:000311921600002,ISI:000301984600017}. 
The EChemDID method introduced in the next subsection, describes the process of equilibration of the applied electrochemical potential in non-simply-connected 
structure of each contact accounting for the loss and gain of atoms during electro-dissolution and metallization.

\subsection{EChemDID equations for electrochemical potential equilibration}

\begin{figure}
  \centering
  \includegraphics[width=0.45\textwidth]{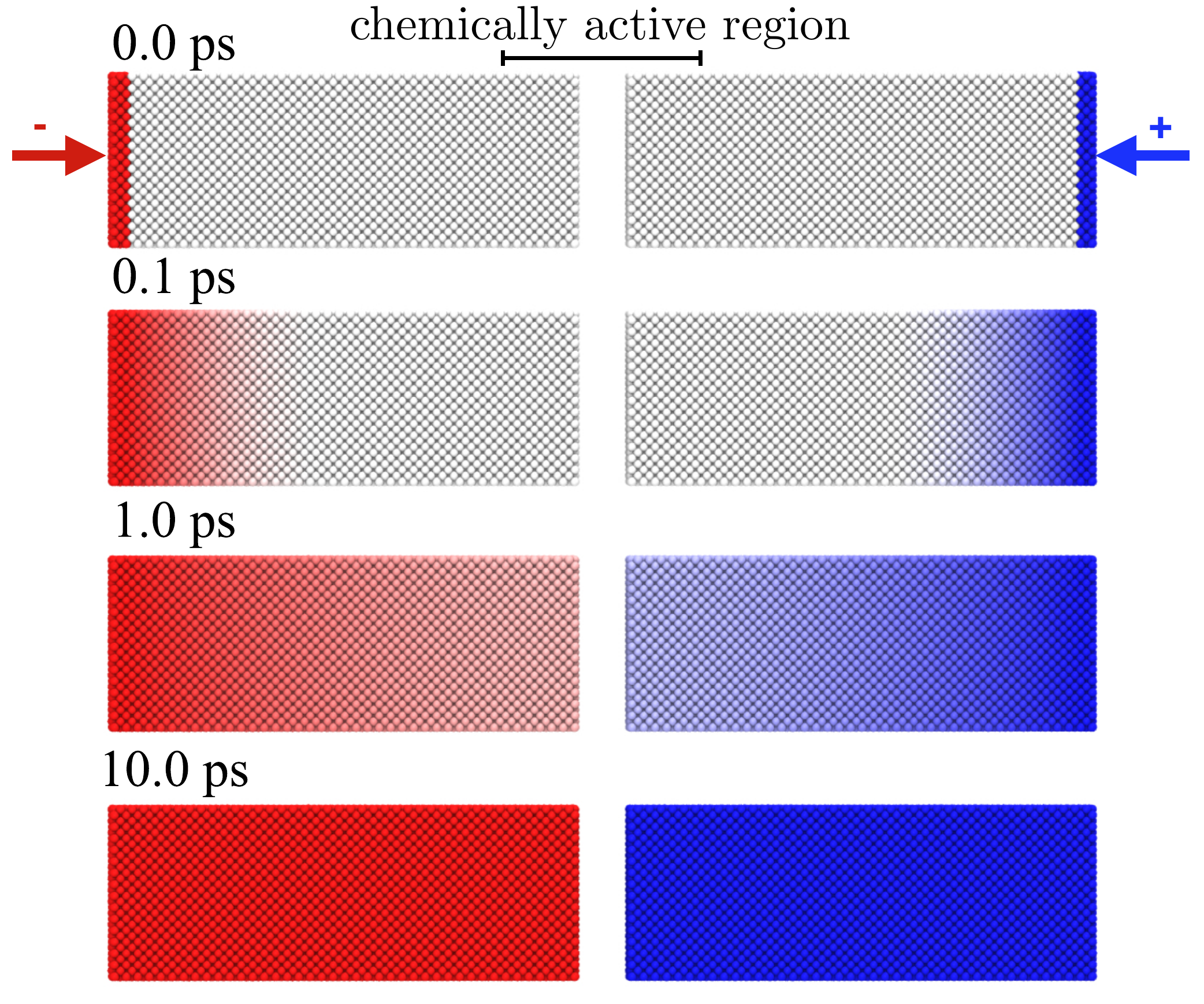}
  \caption{Propagation and equilibration of the chemical potential computed with the EChemDID method.
  The arrows on the top snapshot point toward thin layers of atoms where we apply the voltage $\Phi_0=8$ V. 
  The colors represent the local electrochemical potential $\Phi_i(t)$ ranging from $\chi_0-\Phi_0/2$ (red) to $\chi_0+\Phi_0/2$ (blue) and we use the
  fictitious diffusion coefficient $k = 6$ \AA$^2$/fs (Eq. \ref{eq:lap}).}
  \label{fig:fig1}
\end{figure}

The problem we need to address now is how the external potential applied to a group of atoms in the electrodes propagates and equilibrates throughout
the metallic structure given that their constituent atoms change as electrochemical reactions occur at the electrodes. We start by assigning an additional 
dynamical variable to each metallic atom that represents the local electrochemical potential in their vicinity: $\Phi_i(t)$. 
This is done in the same spirit as the local electronic temperature in two-temperature models designed to capture the thermal transport role of conduction electrons in 
metals~\cite{ISI:A1993LR90100019,ISI:000340713700009} or to represent the temperature of internal degrees of freedom in coarse 
grain simulations~\cite{ISI:000226308000036}. We consider that the electrochemical potential is externally applied to a pre-determined set of atoms located away from the 
chemically active regions, see Figure \ref{fig:fig1} (at 0 ps), and the electrochemical potential of these atoms remain constant throughout the simulation. 
We now propose a dynamical equation for the temporal evolution of the electrochemical potential of all atoms in the system.

The voltage applied to a group of atoms within the metallic electrode propagates through it at the speed of light following Maxwell's wave equations~\cite{jackson91} 
until the entire metal is at the same electrochemical potential. We are not interested in the actual dynamics of this voltage equilibration but on the much slower 
electrochemical reactions; thus we use fictitious diffusive dynamics to describe the equilibration process. We assume that the voltage within the metal follows:

\begin{equation}
\dot{\Phi}=k\nabla^2\Phi
\label{eq:diffusion}
\end{equation}

where the dot denotes time derivative, $\nabla^2$ is the Laplacian operator and $k$ is an effective diffusivity. If a voltage difference is applied to electrodes that are not 
connected to each other, the solution of Eq. \ref{eq:diffusion} will result in the voltage being equilibrated within each electrode at a timescale that depends on the value of $k$. 
Furthermore, we will assume that if an atom becomes detached from the electrode its electronegativity should go back to its atomic value $\chi_i^0$; that is, the external 
electrochemical potential $\Phi$ should go to zero.
The numerical solution of Eq. \ref{eq:diffusion} is performed on-the-fly during the MD simulation using atoms as a grid and a local weighting function as in 
the eleDID method for thermal transport~\cite{ISI:000340713700009}:

\begin{equation}
\dot{\Phi}_i(t)=\sum_{j\neq i}\frac{\Phi_i(t)-\Phi_j(t)}{|R_{ij}|^2}w(R_{ij}) - \eta F(W_i) \Phi_i
\label{eq:lap}
\end{equation}

The first term on the RHS solves the diffusion equation (Eq. \ref{eq:diffusion}); $R_{ij}=R_i-R_j$ denotes the distance between atoms $i$ and $j$ and $w(R)$ is a  
local weighting function defined below. The second term in the RHS relaxes the electronegativity of atoms detached from 
the electrodes to their atomic value. $\eta$ is the relaxation rate, $F(W)$ is a switching function that turns on the relaxation as an atom 
becomes detached from other metallic atoms and $W_i$ is the {\it total metallic coordination} of atom $i$. This external
electrochemical potential is then added to the atomic electronegativity $\chi_i^*(t) = \chi_i^0 + \Phi_i (t)$ and $\chi_i^*(t)$
is used for charge equilibration.

The local weighting function takes the form:

  \begin{eqnarray}
  w(R_{ij}) =
  \begin{cases}
     \mathcal{N} \left[1-\left(\frac{R_{ij}}{R_C}\right)^2\right]^2 & \quad \text{if $R<R_C$}, \\
     0 & \quad \text{otherwise}.\
     \end{cases}
     \label{eq:weight}
  \end{eqnarray}
  
with $\mathcal{N}$ a normalization constant. The range of this weighting function ($R_C$) represents a critical separation distance below which
two atoms are considered as part of the same metallic cluster and, thus, be at the same electrochemical potential.
We define the normalization factor $\mathcal{N}$ from a reference lattice (in $D$ dimensions, with neighbors at distance $R_n$) as: 

\begin{equation}
\mathcal{N} = \frac{2DN_{at}}{\sum_i W_i}
\end{equation}

$N_{at}$ is the total number of atoms and $W_i$ represents the total metallic coordination of atom $i$ in the reference
lattice $W_i=\sum_{j\neq i}w(R_{ij})$. The last term in Eq. \ref{eq:lap} relaxes the external potential of isolated atoms to
zero. The switching function $F(W_i)$ increases from zero to one as the total metallic coordination decreases to zero as:

  \begin{eqnarray}
  F(W) =
  \begin{cases}
     \left[1-\left(\frac{W}{W_0}\right)^2\right]^2 & \quad \text{if $W<W_0$}, \\
     0 & \quad \text{otherwise}.\
     \end{cases}
  \end{eqnarray}
  
$W_0$ represents a critical metallic coordination below which an atom will evolve its electronegativity towards the equilibrium value.
We take this value as that of the weighting function $w$ when two atoms are separated by a distance close to but smaller than
the cutoff $R_C$ (Eq. \ref{eq:weight}). In practice, we take this value as $w(0.99 R_C)$.

Figure \ref{fig:fig1} exemplifies the equilibration of the chemical potential in a two electrodes setup. Each electrode consists of Cu atoms forming a perfect
fcc lattice and an applied external electrochemical potential of $\pm 4$ V is applied to the regions shown at time $t=0$ ps. Solving the EChemDID
equations (with atoms fixed in space in this first example) shows the equilibration of the voltage in each of the contacts. 

The following Sections demonstrate the coupling of EChemDID with charge equilibration that generates electrostatics and with ionic dynamics to simulate electrochemical
reactions.

\subsection{EChemDID and electronic transport: estimating current densities}

While the main purpose of the method is to equilibrate the electrochemical potential in metallic structures for electrochemical simulations, an interesting 
by-product of the approach is that Eqs. \ref{eq:diffusion} and \ref{eq:lap} ($\eta=0$) can be used to estimate electrical currents when a potential difference is applied 
across a metallic system. Assuming diffusive transport, a combination of Ohm's law $\vec{\nabla}\Phi=\rho \vec{j}$ (where $\vec{j}$ is the electrical current 
and $\rho$ the electrical resistivity) with the continuity  equation $\nabla^2\Phi=\rho \vec{\nabla}\vec{j} = \rho\dot{q}$ (where $q$ is the electrical charge density) 
shows that the time derivative in the electrochemical potential in EChemDID Eq. \ref{eq:diffusion} is proportional to the time derivative of the free charge density
as $\dot{\Phi}=k\rho\dot{q}$. 
Let's consider a continuous region $\Omega$ of a metallic wire of cross sectional area $A$ sandwiched between the regions $\Omega_{-1}$ and $\Omega_{+1}$, 
the time derivative of the total charge $\dot{Q}_\Omega$  in $\Omega$ can be expressed as:

\begin{equation}
\dot{Q}_\Omega=\int_\Omega\dot{q}_\Omega d\Omega=\sum_{i\in \Omega}\frac{\nabla^2\Phi_i}{\rho}\Omega_i
\end{equation}

with $\Omega_i$ an atomic volume. From Eq. \ref{eq:lap} (assuming $\eta=0$) $\dot{Q}_\Omega$ can be written as the sum of two terms representing the current flowing to/from
$\Omega_{-1}$ and $\Omega_{+1}$:

\begin{equation}
\dot{Q}_\Omega=\frac{1}{\rho}\left[\sum_{i\in \Omega}\sum_{j\in \Omega_{-1}}\frac{\Phi_i(t)-\Phi_j(t)}{|R_{ij}|}w(R_{ij})\Omega_i+\sum_{i\in \Omega}\sum_{j\in \Omega_{+1}}\frac{\Phi_i(t)-\Phi_j(t)}{|R_{ij}|}w(R_{ij})\Omega_i\right]
\end{equation}
 
Finally, under steady state, the total current density from the EChemDID calculation can be obtained as:

\begin{equation}
j=\frac{1}{\rho A}\sum_{i\in \Omega}\sum_{j \in \Omega_{-1}}\frac{\Phi_i(t)-\Phi_j(t)}{|R_{ij}|^2}w(R_{ij})=\frac{1}{\rho A}\sum_{i\in \Omega}\sum_{j \in \Omega_{+1}}\frac{\Phi_i(t)-\Phi_j(t)}{|R_{ij}|^2}w(R_{ij})
\label{eq:current}
\end{equation}

The equality of the currents coming in and out of the region of interest offers a simple verification of the numerical implementation.

\subsection{EChemDID implementation}

The EChemDID method has been implemented as an external LAMMPS~\cite{Plimpton19951} user-package~\cite{EChemDID} including a LAMMPS ``compute'' that solves the Laplacian in Eq. \ref{eq:lap}
and a ``fix'' that integrates the voltage diffusion in time (Eq. \ref{eq:diffusion}). The whole implementation is consistent with the parallel scheme employed in LAMMPS.
The modifications are independent to the actual LAMMPS program, only minor modifications to the reax/c package~\cite{ISI:000303221300004} allow switching from 
regular QEq (with constant electronegativity) to dynamical electronegativity in EChemDID. 
In the following MD simulations, the interactions between atoms are described by ReaxFF for Cu-SiO$_2$ from the Ref.~\cite{NicoECM}. 
While EChemDID has been fully implemented following the model presented in the previous section, a simplified version will be applied in the following ignoring the 
relaxation term and we will refer to the numerical solution of the Laplacian as Eq. \ref{eq:lap} ($\eta=0$).
This last approximation assumes that the particles retain memory of their previous electronegativity.

\section{Implementation verification and EChemDID validation}

We first verify our implementation of the EChemDID equilibration using atoms as a grid by analyzing the 1D propagation of the electrochemical potential 
through a metallic electrode. We then validate the approach by computing the electric fields generated by the atomic charges induced by the difference
in potential between two electrodes.

\subsection{Electrochemical potential equilibration and electrostatics}

We start by verifying the EChemDID equilibration of voltage by comparing the numerical solution with an analytical one in a 1D transport
over a contact of length $L$ with one end set to +4 V. The initial voltage through the sample is zero and the analytical solution for the diffusion 
equation leads to voltage profiles as a function of time given by:

\begin{equation}
\Phi(x,t) = \chi_0+\frac{\Phi_0}{2}\left(1-\frac{x}{L}\right)-\frac{\Phi_0}{\pi}\sum_n^{+\infty}\frac{1}{n}e^{-\lambda_n^2t}\sin{\frac{n\pi}{L}x}
\label{eq:solution}
\end{equation}

with $\lambda_n=n\sqrt{k}\pi/L$.

\begin{figure}
  \centering
  \includegraphics[width=0.45\textwidth]{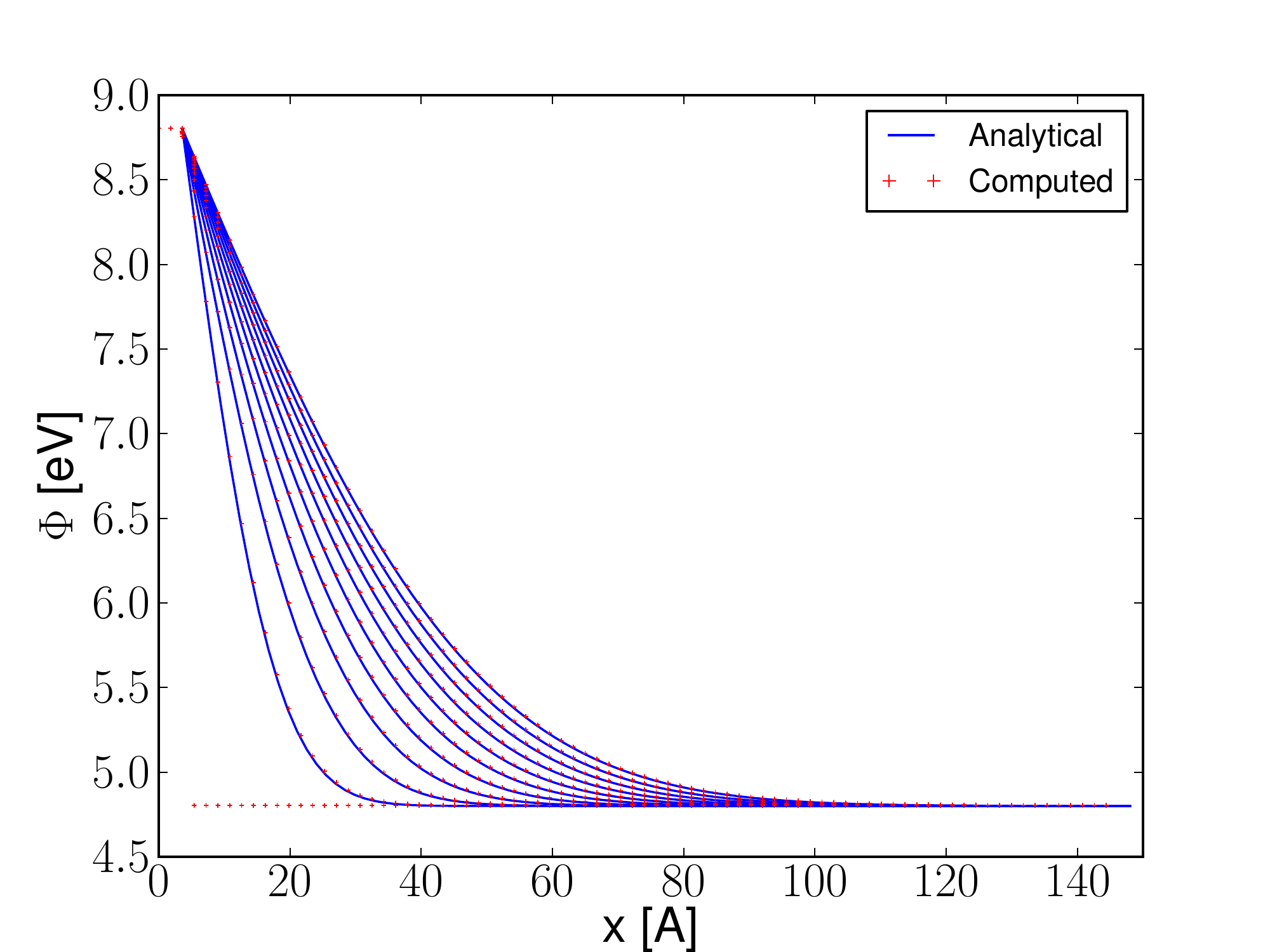}
  \caption{Propagation of voltage along a copper contact computed with EChemDID (Eq. \ref{eq:lap}) compared
  to the analytical solution (Eq. \ref{eq:solution}).
  The chemical potential profile is represented every 10 fs from 0 to 100 fs ($k$ = 6 \AA$^2$/fs and $\Phi_0/2$=+4 V).}
  \label{fig:fig2}
\end{figure}

Figure \ref{fig:fig2} compares the numerical solution (red circles) with the analytical one (lines) at different times; the EChemDID numerical
integration uses a timestep of 0.05 fs.
The excellent agreement verifies our numerical solution of the equation Eq. \ref{eq:lap}. 
For this example, the Cu atoms are fixed in space; coupling Eq. \ref{eq:lap} with molecular dynamics will be demonstrated in Section \ref{sec:mdsim}. 

\begin{figure}
  \centering
  \includegraphics[width=0.9\textwidth]{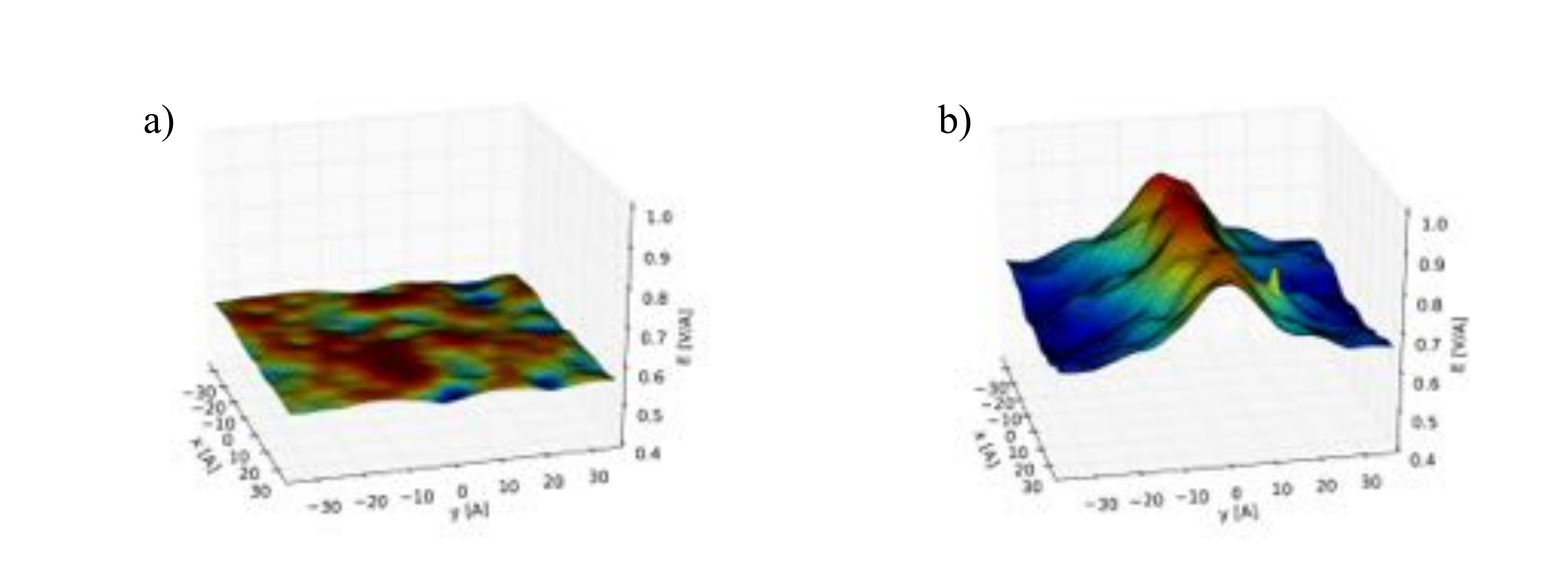}
  \caption{Maps of the electric field between two electrodes under a potential of 8 V: a) the electric field generated by two perfectly flat electrodes (separated by a 
  distance $d = 1.37$ nm, resulting in the voltage $\Phi_0 \approx 7.6$ V); b) the electric field generated by a triangular shaped electrode versus a perfectly flat electrode (separated 
  by a tip-to-flat distance $d = 0.98$ nm, resulting in the tip-voltage $\Phi_0 \approx 8.3$ V) highlighting the effect of field enhancement at the tip of the patterned electrode.}
  \label{fig:fig3}
\end{figure}

Having demonstrated that EChemDID equilibrates the electrochemical potential within metallic electrodes we now show how changing the atomic electronegativity 
affects the QEq-derived atomic charges and, thus, creates electric fields. We do this first for a simple parallel plates capacitor setup separated by vacuum. 
The application of an electrochemical potential $\Phi_0$ between infinite, parallel electrodes separated by vacuum of thickness $d$ should generate a constant 
electric field between them equal to $\Phi_0/d$ and a surface charge density equal to $\sigma=\Phi_0/d\epsilon_0$ (with $\epsilon_0$ the relative permittivity).
After solving the EChemDID equations and performing charge equilibration we compute the magnitude of the electric field between the electrodes using a probe 
charge that scans space. The resulting force $F$ on the probe atom of charge $q$ is proportional to the electric field as $F=qE$. Figure \ref{fig:fig3} shows 2D maps 
of the resulting electric field generated by two types of electrodes. The electric field created by the flat electrodes represented in \ref{fig:fig3}a is spatially constant 
whereas the roughness induced by the triangular electrode showed in \ref{fig:fig3}b results in electric field enhancement at the tip of the electrode.
In both cases depicted in Figure \ref{fig:fig3}, the amplitude of the electric field agrees with the applied voltage with approximately $5\%$ of error.

\subsection{Driving force for electrochemistry}

\begin{figure}
  \centering
  \includegraphics[width=0.45\textwidth]{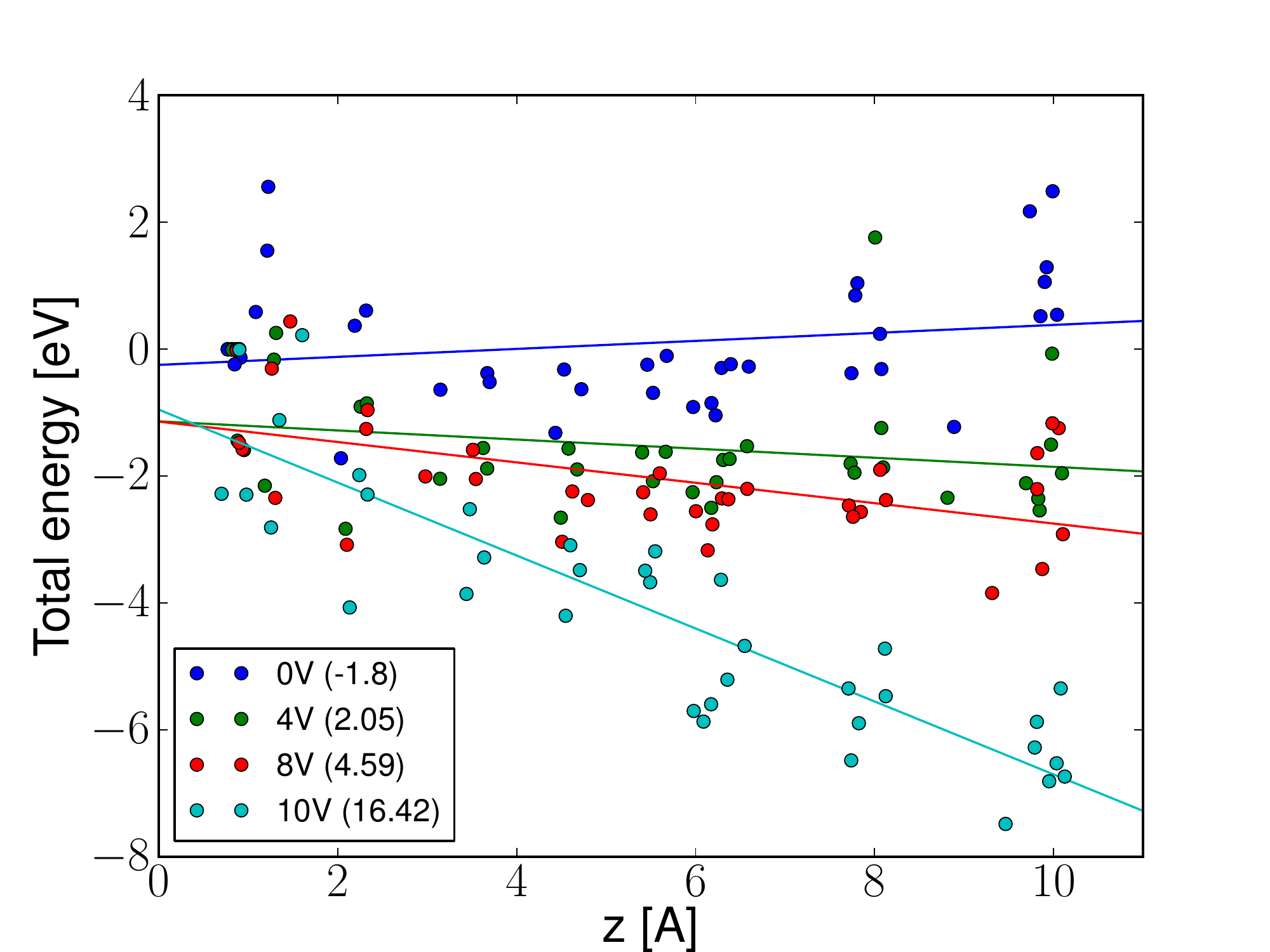}
  \caption{Energy profile of a system composed by a copper probe atom (located in $z$) dissolved in SiO$_2$ sandwiched between two copper electrodes at different voltages.
  The actual voltage applied to the electrodes is shown in the legend and the effective voltage is shown in parenthesis.}
  \label{fig:fig4}
\end{figure}

In order to quantify the driving force for electrochemistry we compare the energy of a series of atomic configurations under different voltages.
We move a probe consisting of a Cu atom through a 1nm-thick layer of crystalline silicon dioxide sandwiched between two copper electrodes.
At each location of the probe atom, we relax the structure via energy minimization and plot in Figure \ref{fig:fig4} the resulting energy landscape.
The voltage applied between the parallel metallic electrodes acts on the ion dissolved in the dielectric and the average slope of the energy landscape 
is given by: $dE/dz = q\Phi_0/d$ where $d$=1 nm is the separation between electrodes and $q$=0.35 e the (averaged) charge of the dissolved ion. 
Therefore, we can extract an effective voltage from the total energies as a function of ion position obtained from the simulations. 
Figure \ref{fig:fig4} compares the extracted effective (in parenthesis) to the applied voltages.
We find a reasonable agreement between the two considering the large fluctuations originating from the rugged underlying atomistic energy landscape.

\section{Simulating electrochemical reactions using EChemDID}
\label{sec:mdsim}

We now focus on full EChemDID simulations where Eq. \ref{eq:lap} ($\eta=0$) is solved together with a reactive MD simulation. To demonstrate the approach we simulate the 
operation of nanoscale ECM cells of interest in nanoelectronics~\cite{ISI:000268309100006}. These resistance-switching devices consist of an electro-active electrode 
(typically Cu or Ag) 
and an electro-inactive one (often W or Pt) separated by a solid electrolyte (amorphous SiO$_2$ in our case). When a voltage of the appropriate polarity is applied, 
the atoms in the active electrode dissolve into the electrolyte and are field-driven towards the inactive electrode. Eventually, this process leads to the formation of a 
metallic conductive filament between the two electrodes, with the subsequent decrease in cell resistance.

\subsection{ECM simulation details}

The simulation cell is composed by two metallic electrodes separated by amorphous SiO$_2$ generated following an annealing procedure described in 
Ref.~\cite{ISI:000290729400009} and similar to the setup presented in Ref.~\cite{NicoECM}. The active electrode is made of Cu and so is the inactive one,
which is described as a rigid body to capture its electrochemical inertness.
In order to mimic the roughness of real interfaces we randomly deleted 50 \% of the Cu atoms part of the first layer at the interfaces. The structure is then 
relaxed via energy minimization and equilibrated using isobaric/isothermal MD for 50 ps. The ECM cell is 4$\times$4 nm$^2$ in cross section where we apply 
periodic boundary conditions and the separation between electrodes is approximately 1.0 nm. 

To simulate the operation of the cell, we impose a constant potential $\pm \Phi_0/2$ to a thin layer of atoms ($\sim$0.2 nm) at 
each boundary of the electrodes (in the direction perpendicular to the interfaces) resulting in a voltage of 8 V, characteristic during the forming phase of this type of 
device~\cite{ISI:000298673500008}. 
The MD equations of motion are solved with a timestep of 0.5 fs within the canonical ensemble at 300 K. The voltage diffusion equation (Eq. \ref{eq:diffusion}) is 
integrated 10 times per MD timestep (for numerical stability) and  the diffusion constant is chosen to be equal to 4 \AA$^2$/fs with a cutoff distance $R_C=$0.4 
nm (Eq. \ref{eq:lap}). For the purpose of demonstration, we use EChemDID with we the coefficient $\eta$ set to 0; a movie of the ECM simulation is included 
in the Supplementary Material.

\subsection{Resistance Switching}

\begin{figure}
  \centering
  \includegraphics[width=0.9\textwidth]{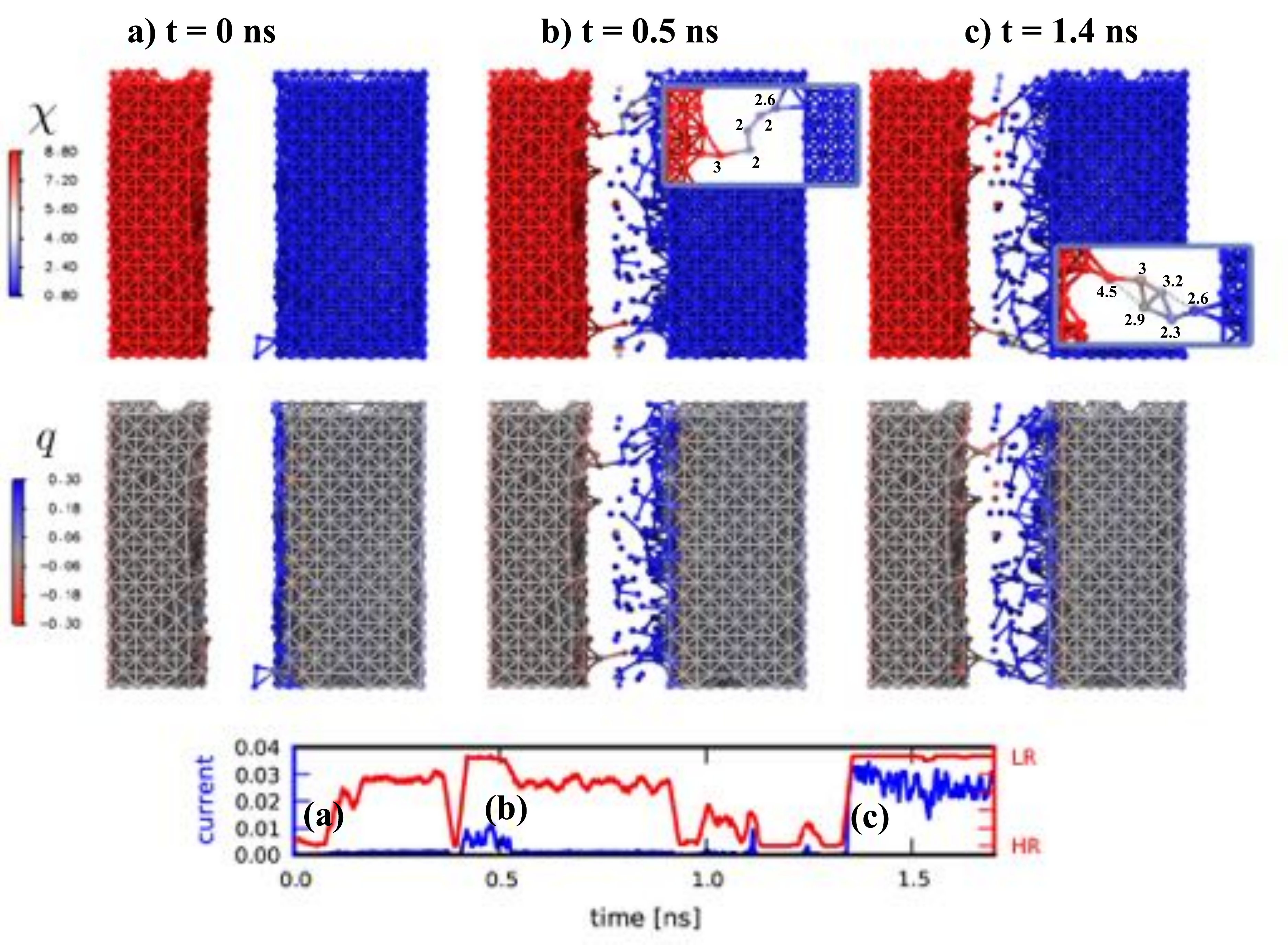}
  \caption{Snapshots of the ECM cell at various times during switching and corresponding switching state of the device computed from Eq. \ref{eq:current} (blue, in arbitrary units)
  or by a distance-based cluster analysis (red, HR/LR = high/low resistance state). 
  The colors on the top panels represent the electronegativity on each atom ranging from 0.8 V (blue) to 8.8 V (red).
  The bottom panels show partial atomic charges ranging from -0.3e (red) to +0.3e (blue). The a-SiO$_2$ between the electrodes has been hidden for clarity.
  We show on the insets in $b$ and $c$ zoomed-in representations of the filament and the local (time-averaged) coordination number of the constituent atoms.
  The coordination numbers in $b$ have been time-averaged from 0.4 to 0.5 ns and the numbers in $c$ have been averaged between 1.4 to 1.6 ns.}
  \label{fig:fig5}
\end{figure}

Figure \ref{fig:fig5} shows atomistic snapshots at various times during the switching of the ECM cell, for clarity only the copper atoms are shown; the bottom panel shows the 
time evolution of the corresponding switching state of the cell. The color on the top row of snapshots denotes the local electronegativity (blue is 
$\chi_0$-4 V and red $\chi_0$+4 V) and the bottom row uses color to denote partial atomic charges. We see that QEq correctly localizes charges at the interfaces between the copper 
electrodes and the amorphous silica with positive ions on the active electrode and negative at the inactive one. The low electronegativity in the active electrode leads 
to positive partial charges which facilitates its dissolution into the dielectric; this can be formally thought of as an oxidation reaction: 
Cu $\rightarrow$ Cu$^{\delta +}+\delta e^-$. 
Similarly, the electronegativity of the dissolved copper ions increases as they become closer to the inactive electrode (within the cutoff R$_C$), lowering their partial 
charge until they bond to the electrode and become metallic following chemical reduction, formally: Cu$^{\delta +}+\delta e^-\rightarrow$ Cu. Throughout the simulation, 
EChemDID maintains constant electrochemical potential within each electrode even as their morphology changes dramatically during the operation of the device. 
As shown above, the charged electrodes generate an electric field throughout the dielectric that  acts as the driving force for the diffusion of copper ions inside the 
solid electrolyte from the active toward the inactive electrode during the forming or programing phase.

EChemDID also provides insightful information regarding the formation of a conducting filament. To study this process we use two approaches: 
i) a distance-based cluster analysis to determine whether the two electrodes are connected by a Cu filament as was done in Ref.~\cite{NicoECM} and ii) 
using the current computed from EChemDID Eq. \ref{eq:current}. The distance-based cluster analysis uses a cutoff of 0.4 nm and we assign a value of
1 to the signal when the two electrodes are bridged (low resistivity state: LR), 0 otherwise (high resistivity state: HR); the red line in Figure \ref{fig:fig5} (bottom scale) shows a running average (with a 
time-window of 25 ps) 
of the resulting signal for clarity. Therefore, a signal between LR and HR corresponds to switching states where the cell oscillates between ON and OFF
within the time-window of the average. The current represented on the bottom scale of Figure \ref{fig:fig5} (blue) has also been averaged with a 
time-window of 5 ps.

From the cluster analysis, we see that a filament was established for a short duration just before 0.5 ns and a more stable
one forms at time 1.4 ns; this second filament remains for the duration of the run. The atomic configuration of the first bridge consists of a linear 
single-atom chain (inset $b$) and bridges the two electrodes for approximately 0.1 ns. As in our prior work~\cite{NicoECM} single-atom chains are
metastable and do not survive long timescales. The second, more stable, filament (inset $c$) is slightly thicker with a zig-zag shape.
We show in the insets of Figure \ref{fig:fig5} ($b$ and $c$) the time-averaged coordination number of the atoms in the filament. While the atoms in 
the first bridge exhibit an averaged coordination number between 2 and 3, typical of a linear configuration, the atoms in the zig-zag chain have 
higher coordination number suggesting a stronger connection.

The current computed via Eq. \ref{eq:current} correlates with the distance-based cluster analysis however additional informations can be extracted 
from its amplitude. Contrary to the binary state of the cluster analysis, the amplitude of the current corroborates the strength of the connection, hence 
the quantity of current that can flow through the filament. We see on Figure \ref{fig:fig5}$c$ a peak in the current approximately 3 times larger than 
in $b$ in agreement with the inset $c$ showing a zig-zag configuration of the bridge. We demonstrate a direct connection between the atomic structure 
of the filament and the current leaving and entering the electrodes computed from the dynamical equilibration of the EChemDID method. 
This last analysis is the qualitative {\it in silico} equivalent to atomically-controlled quantum conductance experiments proposed recently~\cite{ISI:000304320300020}.

\section{Discussion and conclusions}

In summary, we introduced a novel approach to describe electrochemical reaction using reactive molecular dynamics simulations.
EChemDID describes the equilibration of electrochemical potential throughout non-simply connected metallic structures and this
potential shifts the energetics used to obtain atomic charges in charge equilibration. Thus, the application of
an external potential leads to atomic charges that can drive electrochemical process. We demonstrated that the driving
force for electrochemistry is accurately described by EChemDID. In addition, the electrochemical potential equilibration is
performed on-the-fly during the MD simulation and thus adjusts to changes of composition and topology of the electrodes.
As atoms dissolve into the electrolyte their electronegativity evolves back to its original value or keeps memory of previous voltage.
The implementation in LAMMPS is particularly straightforward and the method can be easily extended to more sophisticated charge 
equilibration methods.

EChemDID is not without limitations. The ions dissolved in the electrolyte could trap electrons and change their charge
state; such processes are currently ignored in EChemDID. Recently, M\"{u}ser and Dapp proposed a method to handle integer
charge states and used it to explore redox reactions in a model battery; such an approach could be incorporated into EChemDID. 
In addition, while EChemDID enables an estimation of electronic currents, they do not affect atomic dynamics; processes like 
Joule heating and electromigration are ignored in this first version of the model.

The demonstration of the use of EChemDID to simulate the resistance switching in ECM cells shows the power of the approach which
is generally applicable with reactive and non-reactive force fields (as long as dynamical charges are computed with QEq) 
and can be used to study electrochemical processes with atomistic detail. 
There is growing interest in {\it ab initio} descriptions of electrochemistry~\cite{ISI:000315141600017}, and EChemDID can help bridge such methods toward full device simulations.

\FloatBarrier
\subsection*{Acknowledgement}
This work was supported by the FAME Center, one of six centers of STARnet, a Semiconductor Research Corporation program sponsored by MARCO and DARPA.
We would like to express our gratitude to Steve Plimpton and Aidan Thompson for the interesting discussions related to the software LAMMPS.
Finally, we thank nanoHUB.org and Purdue for the computational resources.
 
\clearpage

\end{document}